\documentclass[useAMS,usenatbib]{mn2e}

\usepackage{amsmath,amssymb,amsbsy,bm}

\providecommand{\adsurl}[1]{\href{#1}{ADS}}

\usepackage{graphicx}
\usepackage[usenames]{color}

\usepackage[hypertex]{hyperref}

\definecolor{myred}{rgb}{0.85,0.08,0}
\definecolor{mydb}{rgb}{0,0.08,0.8}

\newcommand{\beq}{\begin{equation}}
\newcommand{\eeq}{\end{equation}}
\newcommand{\bea}{\begin{eqnarray}}
\newcommand{\eea}{\end{eqnarray}}
\def\bi{\begin{itemize}}
\def\ei{\end{itemize}}

\title[CMB TE and PGWs]{CMB Temperature Polarization Correlation and Primordial Gravitational Waves}

\author[A.G. Polnarev, N.J. Miller, and B.G. Keating]{A.G. Polnarev$^{1}$\thanks{E-mail:A.G.Polnarev@qmul.ac.uk}, 
N.J. Miller$^{2}$\thanks{E-mail:nmiller@physics.ucsd.edu} and B.G. Keating$^{2}$\thanks{E-mail:bkeating@ucsd.edu} \\
$^{1}$Queen Mary, University of London, Mile End Road, London, United Kingdom \\
$^{2}$University of California, San Diego, Mail Code 0424, La Jolla, CA, 92093-0424, United States of America}

\pagerange{\pageref{firstpage}--\pageref{lastpage}} \pubyear{2007}

\def\LaTeX{L\kern-.36em\raise.3ex\hbox{a}\kern-.15em
    T\kern-.1667em\lower.7ex\hbox{E}\kern-.125emX}

\begin{document}

\label{firstpage}

\maketitle

\begin{abstract}
We examine the use of the CMB's TE cross correlation power spectrum as a complementary test to detect primordial gravitational 
waves (PGWs). The first method used is based on the determination of the lowest multipole, $\ell_0$, 
where the TE power spectrum, $C_{\ell}^{TE}$, first changes sign. The second method uses Wiener filtering on the CMB TE data to remove the 
density perturbations contribution to the TE power spectrum. In principle this leaves only the contribution of PGWs. We examine 
two toy experiments (one ideal and another more realistic) to see their ability to constrain PGWs using the TE power spectrum alone. 
We found that an ideal experiment, one limited only by cosmic variance, can detect PGWs with a ratio of tensor to scalar metric perturbation power spectra 
$r=0.3$ at $99.9\%$ confidence level using only the TE correlation. 
This value is comparable with current constraints obtained by WMAP based on the $2\sigma$ upper limits to the B-mode amplitude. We 
demonstrate that to measure PGWs by their contribution to the TE cross correlation power spectrum in a 
realistic ground based experiment when real instrumental noise is taken into account, the 
tensor-to-scalar ratio, $r$, should be approximately three times larger.
\end{abstract}


%\catchline{}{}{}{}{}
%
%%%%%%%%%%%%%%%%%%%%%%%%%%%%%%%%%%%%%%%%%%%%%%%%%%%%%%%%%%%%%%%%%%%%

\section{Introduction}

Primordial gravitational waves (PGWs) polarize the cosmic microwave background (CMB) (see for example \cite{basko, Polnarev85, 
Crittenden93, Frewin94, colefrewinpolnarev95, kamionkowski97, 
Seljak97Pol, selzal97, Kamionkowski98, baskaran06, Keating2006}). Current 
experiments are using the polarization of the CMB to search for this PGW background (\cite{CloverIntro, BowdenOpt, BicepPaper}. This polarization can 
be used as a direct test of inflation. An alternative probe of the inflationary epoch which does not use the PGW background was 
studied by (\cite{sperzal97}). This probe 
was used in recent analyses by the WMAP team (\cite{peiris03}) to provide plausibility for the inflationary paradigm. This paper presents a test 
similar in spirit to that of \cite{sperzal97}. 

CMB polarization can be separated into two independent components: E-mode (grad) polarization and 
B-mode (curl) polarization. B-mode polarization can only be generated by PGWs (see for example \cite{Seljak97Pol, selzal97, Kamionkowski98}), therefore 
most CMB polarization experiments which are searching for evidence of PGWs focus on measuring the 
BB power spectrum. However the TE cross correlation power spectrum offers another method to detect PGWs (\cite{crittenden94}). 
The TE power spectrum is two orders of magnitude 
larger than the BB power spectrum and it was suggested that it may therefore be easier to detect gravitational waves in the 
TE power spectrum (\cite{baskaran06, grishchuk07}).

In this paper we first discuss the method of detection of PGWs by measuring the TE power spectrum for low $\ell$. 
This method, originally proposed in \cite{baskaran06}, is based on a measurement of $\ell_0$, the multipole where the TE power spectrum first changes sign. 
Hereafter we will call this method ``the 
zero multipole method''. The TE power spectrum due to density perturbations is positive on large scales, 
corresponding to $\ell < \ell_0$, changes sign at $\ell = \ell_0$, and then oscillates for $\ell > \ell_0$, while for PGWs the TE power 
spectrum must be negative for small $\ell$ and 
then also oscillates for larger $\ell$. The current best set of cosmological 
parameters, obtained in \cite{Spergel2007WMAP}, gives, in the absence of PGWs, $\ell_0 = 53$. Therefore, the measurement of the difference 
between the multipole 
number, $\ell_0$, where the TE power spectrum changes sign, and $\ell=53$ is the way to detect PGWs.  We will then consider an alternative 
method based on Wiener filtering, removing the contribution to the TE power spectrum due to density perturbations. Since the TE 
power spectrum due to PGWs is megative on large scales a test of negativity of the resulting TE power spectrum is a test of PGWs. In this 
paper, we present an analysis of both of these methods, based on Monte Carlo simulations. 

At the present time, the main priority and the main challenge in CMB polarization observations is the detection of the 
PGW background via the BB power spectrum. In connection with BB experiments, the methods based on the TE cross correlation can be considered as very useful 
auxiliary measurements of PGWs because
systematic effects in TE measurements are not degenerate with those in BB measurements. For example, T/B leakage or even E/B leakage could swamp a
detection of BB, whereas T/E leakage would be small and well
controlled (see \cite{meir07}). These BB systematics could falsely indicate a detection of PGWs, but measurements of the TE power spectrum
provide insurance against such a spurious detection. Additionally, galactic foreground contamination affects BB and TE in different ways, which 
enables us to perform powerful cross-checks and subtraction of foregrounds in BB measurements.

Another advantage of TE measurements for experiments which measure a small fraction of the sky, is related to the fact that 
a significant contaminant to the B modes is caused by E/B mixing. This limits the power spectrum of PGWs that 
can be detected (\cite{challinor05}). The E-modes are practicly unaffected by E/B mixing so, in contrast to the BB measurements, the TE power spectrum 
should be nearly the same for both full and partial sky measurements.

The plan of this paper is the following. In Section \ref{physmath}, we introduce the primordial power spectra of scalar (density) and tensor 
(PGW) perturbations (\ref{primspectra}). Then following \cite{crittenden94} and \cite{baskaran06}, we  
explain why the sign of the TE power spectra for scalar and tensor perturbations is opposite for large scales (\ref{TECross}). In 
Section \ref{ell0det}, we describe in more detail the 
zero multipole method for the detection of PGWs. In Section \ref{wienfilt}, we describe the method for detection of PGWs based on 
Wiener filtering along with the statistical tests used and a comparison of the tests. In 
Section \ref{resul}, we present results of numerical Monte Carlo 
simulations for two toy experiments. In the first toy
experiment we neglect instrumental noise and the uncertainties are limited only by cosmic variance (\ref{toy1res}). In 
the second toy experiment, along with cosmic variance, 
we take into account instrumental noise which is comparable to real noise in 
current ground experiments (\ref{toy2res}). For comparison, we also present results of simulations for the two satellite experiments, 
WMAP (\ref{wmapres}) and  Planck (\ref{planckres}). In Section \ref{comptebb}, we compare the the signal-to-noise ratio of the TE 
measurements with those of BB measurements.

\section{TE Cross Correlation} \label{physmath}

The power spectrum of TE correlations is determined by primordial power spectra of scalar and tensor perturbations and time evolution 
of these perturbations during the epoch of recombination.

\subsection{Primordial Power Spectra} \label{primspectra}

The primordial power spectra describing the initial scalar (density) perturbations (denoted by $s$) and tensor (PGW) perturbations 
(denoted by $t$) are (see, for example, \cite{Spergel2007WMAP})

\begin{eqnarray}
	\qquad \qquad \qquad P_s(k) &=& A_s \left( \frac{k}{k_0} \right)^{n_s - 1 + \frac{1}{2} \alpha_s \log \left(k/k_0\right)} \nonumber \\
	P_t(k) &=& A_t \left( \frac{k}{k_0} \right)^{n_t} \label{AtAs},
\end{eqnarray}
where $k_0 = 0.002 $ Mpc$^{-1}$, this value of $k_0$ 
is obtained by fitting of CMB data (\cite{Smith2006}).
The variables $n_s$ and $n_t$ are the scalar and tensor spectral indices respectively. The variable $\alpha_s$ is the running of the 
scalar spectral index. In terms of $A_s$ and $A_t$, the tensor-to-scalar ratio, $r$, is 

\begin{equation}
	r \equiv \frac{A_t}{A_s} = \frac{P_t(k_0)}{P_s(k_0)} \label{req}
\end{equation}

The location of $\ell_0$ is determined by the parameters $n_t$ and $r$. 
In this paper, we do not specify particular cosmological models considering the generation of primordial spectra, $P_s(k)$ and $P_t(k)$, 
which means that for our purposes we consider $n_s$, $n_t$, 
and $r$ as independent parameters. (This is not true if we use some particular cosmological model. For example, in standard inflation 
models, the parameters $n_t$ and $r$ are related by 
the consistency relation, $n_t = -r / 8$ (see, for example, \cite{peiris03}).) In other words, we consider 
all parameters $n_s$, $n_t$, and $r$ as independent except in \ref{toy1res} and \ref{toy2res}, where along with model-independent we 
give also model-dependent constraints on $r$.

\subsection{Opposite Signs of Scalar and Tensor Perturbations to TE Correlation} \label{TECross}

Taking into account that scalar and tensor perturbations are not
correlated, the TE power spectrum is simply a sum of two
TE power spectra for scalar and tensor perturbations
correspondingly.

First, the physical motivation for the difference in the cross correlation 
contributions produced by
scalar and tensor perturbations for small $\ell$ was demonstrated
and physically interpreted for the cross correlation of the Stokes
parameters $T$ and $Q$ in \cite{crittenden94}. For scalar
perturbations the Stokes parameter $Q$ contains only E-modes, hence the TE
correlation is identical with the TQ correlation and is positive for
small $\ell$. As was then emphasized in \cite{baskaran06}, the sign
of the TE correlation for tensor perturbations is negative for small
$\ell$. The simple qualitative physical interpretation of the fact
that the contributions of the TE correlation are different for scalar and tensor
perturbations is the following. For both scalar and tensor
perturbations, the temperature fluctuations, $T(\ell)$, for small
$\ell$ (when oscillations of $T(\ell)$ are absent) are proportional
to the metric perturbations $h$ at the moment of
recombination, while the E-mode fluctuations, $E(\ell)$, are
proportional to $\dot{h}$ at the moment of recombination. Hence, the TE
correlation is proportional to $h \dot{h} \propto \frac{d(h^2)}{d\eta}$.
Taking into account the growth of scalar perturbations and tensor perturbations 
decay, one can see that the contributions to the TE
correlation for scalar and tensor perturbations are opposite.

To understand this in more detail, following
\cite{baskaran06}, we consider the multipole expansion of the TE cross
correlation with coefficients $C_{\ell}^{TE}$. These coefficients
are related to the spherical harmonic expansion coefficients of the
temperature anisotropy and polarization by

\begin{equation}
  C_{\ell}^{TE} = \left\langle a_{T, \ell m} a_{E, \ell m}^{*} \right\rangle
\end{equation}
where the brackets denote averaging over all possible statistical
realizations. The statistical properties of the CMB field in
general, and the TE cross correlation specifically, follow from the
statistical properties of the underlying scalar or tensor metric
field. Assuming gaussianity together with statistical isotropy and
homogeneity, the TE cross correlation takes the form

\begin{equation}
C_{\ell}^{TE} = \int \frac{dk}{k} a_{T,\ell} (k)a_{E,\ell} (k)
\end{equation}
where $a_{T,\ell} (k)$ is the contribution from temperature
perturbation while $a_{E,\ell} (k)$ is the contribution from
E-polarization. The integration over $k$ takes into account the
contribution from all the possible wavenumbers.

It was shown in \cite{baskaran06}

\begin{equation}
  a_{T,\ell} (n) \sim h_k(\eta) \bigg|_{\eta=\eta_{rec}}
\end{equation}

\begin{equation}
  a_{E,\ell} (n) \sim  \frac{d h_k(\eta)}{d\eta} \Bigg|_{\eta=\eta_{rec}}
\end{equation}
where $h_k(\eta)$ is the mode function of the metric perturbation,
and $\eta_{rec}$ is the conformal time at recombination. It follows
that the TE correlation is approximately

\begin{equation}
  C_{\ell}^{TE} \propto \int dn F(\ell,k) \left( \frac{d h_k^2 (\eta)}{d\eta} \right) \bigg|_{\eta=\eta_{rec}} \label{gwte}
\end{equation}
where $F(\ell,k)$ is a strictly positive function which peaks at
$\sim \ell \approx k\left(\eta_{today}-\eta_{rec}\right)$.
Heuristically, the function $F(\ell,k)$ projects the $k$-space onto 
the $\ell$-space. Therefore the sign of
the integral in the RHS of Eq. \ref{gwte} evaluated at around $\ell
\approx k\left(\eta_{today}-\eta_{rec}\right)$ determines the sign
of $C_{\ell}^{TE}$ on large scales.

The adiabatic decrease of the gravitational wave amplitude upon
entering the Hubble radius is preceded by the monotonic descrease of
the gravitational wave mode function $h_k(\eta)$ as a function of
$\eta$. Since $h_k(\eta)$ is decreasing the integral on the RHS of
Eq. \ref{gwte} is negative. The RHS of Eq. \ref{gwte} is negative
for $k\left(\eta_{today}-\eta_{rec}\right) < 90$ since $h_k$ is
decreasing over that range. Therefore, for $\ell < 90$ the
correlation $C_{\ell}^{TE}$ must be negative. For larger $\ell$s,
the $F(\ell,k)$ in Eq. \ref{gwte} and, hence, the TE cross
correlation power spectrum changes sign as a function of $\ell$.

Thus the TE cross correlation, due to density pertrbations, must be
positive at lower $\ell$ (as mentioned above, the TE cross
correlation in absence of PGWs changes sign at $\ell_0 \approx 53$).
If we were able to separate them we could use this signature for
detection of PGWs. However, even without
such separation the presence of PGWs manifests itself in the value
of $\ell_0$, which is the smallest $\ell$ where the total TE
correlation power spectrum (scalar plus tensor) changes its sign.
Thus, the sign of the TE correlation is a very prominent signature
of PGWs. For this reason, in the next section, we investigate the
dependance of $\ell_0$ on $r$, $n_s$, and $n_t$. 

\section{Dependance of $\ell_0$ on Parameters of PGW Power Spectrum} \label{ell0det}

The method of detecting PGWs which implies a calculation of $\ell_0$, where the TE power spectrum first goes to zero, will be 
called hereafter as the zero multipole method. 
We take into account uncertainties in determination of $C_{\ell}$s which are unavoidable in any experiment: 

\begin{eqnarray}
	\qquad (\Delta C_{\ell}^{TE})^2 &=& \frac{1}{(2\ell+1)f_{sky}} \Bigg( (C_{\ell}^{TE})^2 + \nonumber \\
	&& (C_{\ell}^{TT} + N_{\ell}^{TT})(C_{\ell}^{EE} + N_{\ell}^{EE}) \Bigg)
\end{eqnarray}
(see, for example, \cite{dodelsonbook}). Even in an ideal experiment, when we neglect instrumental 
noise ($N_{\ell}=0$) and measure the full sky ($f_{sky}=1$), we still have uncertainties related with cosmic variance (which arises 
from the fact that we have 
only one realization of the sky in CMB measurements) (see, for example, \cite{dodelsonbook}). For a more realistic experiment, 
we take into account noise and partial sky coverage (see Section \ref{resul}). Over small multipole bands it is reasonable to 
approximate the power spectrum as linear. In the 
range $20 \le \ell \le 70$, it seems reasonable to use a linear approximation for $(\ell+1) C_{\ell} / 2\pi$. It seems unlikely that 
in this range any deviations from a linear approximation can be 
larger than mentioned above uncertianties. 

\begin{figure}
    \begin{center}
    	\includegraphics[totalheight=0.25\textheight]{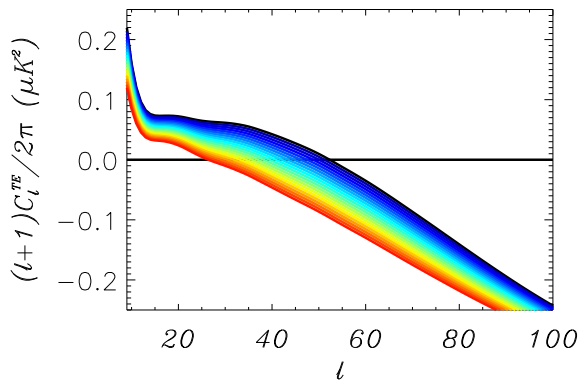}
    	\\
		\caption{\small The TE cross correlation power spectrum for different values of $r$ with $n_t = 0$. The black line is $r=0.0$ 
		and the red line is $r=2.0$. Lines are given for $0<r<2.0$ with a spacing of $\Delta r=0.1$.}
		\label{fig5}
	\end{center}
\end{figure}

Plots of $C^{TE}_{\ell}$ for different values of $r$ are shown in Fig. \ref{fig5} plotted for $n_t=0$. It 
can be seen that a linear fit to the TE power spectrum do well approximates $(\ell+1) C_{\ell}^{TE}/2\pi$ near $\ell_0$.

Thus near $\ell_0$, $(\ell+1)C_{\ell}^{TE}/2\pi$ can be approximated as a line with negative slope $a-b\ell$, where $a$ and $b$ are positive 
real numbers. For any set of experimental data, we can find $a$ and $b$ by applying a least squares fit. The values $a$ and $b$ 
corresponding to the best fit obviously can be used 
for prediction of $\ell_0 = a/b$. This value, $\ell_0$, can then be used to constrain the parameter $r$ under some assumptions 
about spectral indices $n_s$ and $n_t$.

\begin{figure}
    \begin{center}
    	\includegraphics[totalheight=0.25\textheight]{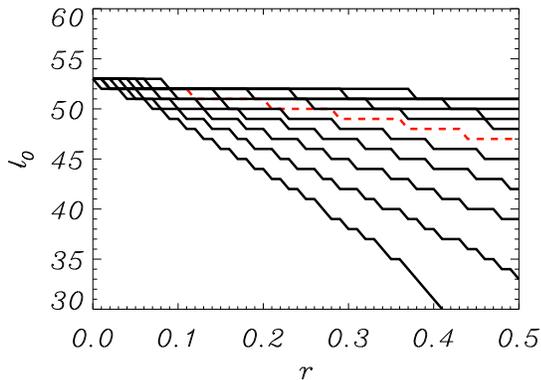}
    	\\
    	\caption{\small Plot of crossover multipole number $\ell_0$ for different values of $n_t$. $n_t = -0.5$ to $0.5$ with spacings of $0.1$. 
		The dashed red line correspond to $n_t=0$.}\label{fig26}
	\end{center}
\end{figure}

\begin{figure}
	\begin{center}
		\includegraphics[totalheight=0.25\textheight]{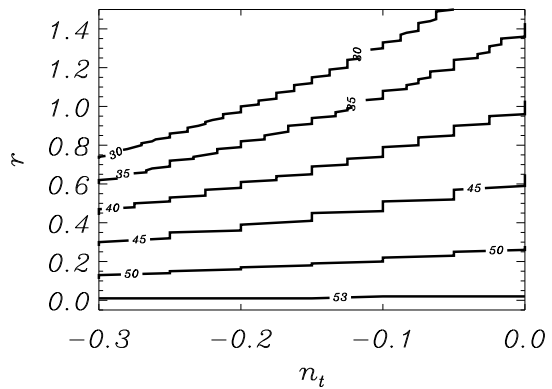}
		\\
		\caption{\small A contour plot of the values of $\ell_0$ for differing values of $r$ and $n_t$.}
		\label{ell0contour}
	\end{center}
\end{figure}

We need to investigate how $\ell_0$ depends on the cosmological parameters $r$, $n_t$, 
$n_s$, $A_s=P_s(k_0)$, and the optical depth to reionization, $\tau$. The value of $\ell_0$ for a standard $\Lambda$CDM cosmology 
described in \cite{Spergel2007WMAP} as a function of 
$n_t$ and $r$ is shown in Fig. \ref{fig26} and \ref{ell0contour}. All power spectra were generated with the code 
CAMB\footnote{see http://camb.info on web} (\cite{Lewis2000}). If $r=0$, $\ell_0 = 53$, while if $r=0.3$ (the WMAP3 upper limit on the 
tensor-to-scalar ratio) and $n_t = 0$, we find that $\ell_0 = 49$.

From Fig. \ref{fig26} and \ref{ell0contour}, one can see that $\ell_0$ decreases with increase of $r$. This effect is more 
pronounced for smaller $n_t$. For example, 
if $r=0.3$, then $\ell_0=52$ for $n_t = +0.5$, 
$\ell_0=49$ for $n_t = 0$, and $\ell_0=38$ for $n_t=-0.5$. The fact that $\ell$ is discrete (the plots are composed of a 
set of step functions) puts limitations on 
using this method for determination of $r$. For example, a value of $r=0.01$ and $r=0$ will most likely correspond to the 
same $\ell_0$ and therefore no matter 
what the sensitivity is this method cannot distinguish between absence of PGWs and PGWs corresponding to such small $r$. 
For $n_t=0$ (the Harrison-Zel'dovich 
scale-free spectrum), $\delta \ell_0 = -1$ corresponds to $\delta r$ of $0.08$. For negative values of $n_t$, $\delta \ell_0 = -1$ 
corresponds to smaller $\delta r$. For $n_t = -0.5$, for example, $\delta \ell_0 = -1$ requires $\delta r \approx 0.02$. 

The effect of variations of the scalar spectral index $n_s$ on $\ell_0$ is opposite: A decrease of $\ell_0$ with increase of 
$r$ is more pronounced for larger $n_s$, however we do not need to worry about it because $n_s$ is well constrained by the 
observations of TT and EE power spectra (see for example \cite{Spergel2007WMAP}) 
along with Ly-$\alpha$ measurements (see for example \cite{lyapaper}). Thus everywhere in this paper we use $n_s = 0.95$ (the 
value given by WMAP3 \cite{Spergel2007WMAP}) with no running of the scalar spectral index, $\alpha_s$. A change of $0.2$ in the running of the 
scalar spectral index has no effect in the value of $\ell_0$ when using $k_0 = 0.002$ Mpc$^{-1}$. Even if it did have an effect on the value of 
$\ell_0$, it can be well constrained by the same observations that constrain $n_s$.

The value of $\ell_0$ does not depend on $A_s$, because for fixed $r$, $A_t$ must change by the same factor as $A_s$ leaving 
$\ell_0$ unchanged, i.e. any rescaling of the primordial power spectra does not change $\ell_0$. The same thing happens 
when one varies the optical depth to reionization, $\tau$. 
The value of $\ell_0$ is in the range where the TE power spectrum for scalar and tensor perturbations depend on $\tau$ in the
same way for instantaneous reionization (they are damped by the factor, $\exp (-2\tau)$, 
since the relevant scales are within the cosmological horizon at the time of reionization (\cite{dodelsonbook}). Thus any variation 
of $\tau$ can be considered just as a rescaling of the TE power 
spectrum which leaves the value of $\ell_0$ unchanged. It is possible for different reionization histories to cause a change in $\ell_0$ as 
shown in \cite{kaplinghat2003}, however we will assume instantaneous reionization for the purpose of this paper.

Thus, even if we cannot separate the contributions of scalar and tensor perturbations to the TE power 
spectra, PGWs still leave their imprint on the value of $\ell_0$.
In the next section, we will consider the possibility of such separation with the help of Wiener filtering.

\section{Wiener Filtering of the TE Cross Correlation Power Spectrum} \label{wienfilt}

Wiener filtering has been used often in the case of CMB data analysis. For example, it was used to combine multi-frequency data in order 
to remove foregrounds and extract the CMB signal from the observed 
data (\cite{Tegmark1996, Bouchet1999}). Here we examine the use of the Wiener filter to subtract the PGW signal 
from the total TE correlation signal. This is done because the Wiener filter reduces the contribution of noise in a total signal by 
comparison with an estimation 
of the desired noiseless signal (\cite{vaseghibook}). In our case, the signal is the one due to PGWs only, and the signal contributed 
by density perturbations is considered to be ``noise''.

The observed signal can be written as

\begin{equation}
	C_{\ell}^{TE} = C_{\ell,s}^{TE} + C_{\ell,t}^{TE} = \left \langle a_{E,\ell m}^* a_{T,\ell m} \right \rangle
\end{equation}
where $s$ and $t$ refer to the contributions to the power spectrum due to scalar and tensor perturbations respectively. The values 
$a_{E,\ell m}$ and $a_{T,\ell m}$ refer to the 
spherical harmonic coefficients of the temperature and polarization maps. In our application to TE correlation, we consider the Wiener 
filter, $W_{TE,\ell}$: 

\begin{equation}
	W_{TE,\ell} = \frac{C_{\ell,t}^{TE}}{C_{\ell}^{TE}} = - \frac{ \left| C_{\ell,t}^{TE} \right |}{C_{\ell}^{TE}}
\end{equation}
The filtered signal, $a_{X,\ell m}^{\prime}$ (for $X = T$ and $E$), is obtained from the measured signal, $a_{X,\ell m}$, as

\begin{equation}
	a^{\prime}_{X, \ell m} =  a_{X, \ell m} W_{TE, \ell}^{1/2}
\end{equation}
In this paper, we assume the Wiener filter is perfect, in the sense that it leaves the signal due to PGWs only. We then get, for the 
filtered multipoles $C_{\ell,filt}^{TE}$,

\begin{eqnarray}
	\qquad \qquad C_{\ell,filt}^{TE} &=& \left \langle a_{T,\ell m}^{\prime *} a_{E,\ell m}^{\prime} \right \rangle \nonumber \\
	&=& W_{TE, \ell} C_{\ell}^{TE} = \frac{C_{\ell,t}^{TE}}{C_{\ell}^{TE}} C_{\ell}^{TE} = C_{\ell,t}^{TE}
\end{eqnarray}

In practice this is not true, because we are trying to 
determine $C_{\ell,t}^{TE}$, which is not known in advance. Nevertheless, the assumption that the Wiener filter is perfect is good as a first 
approximation and illustrates the 
detectability of PGWs with the help of TE correlation measurements.

The filtering can reduce the measured signal to the desired signal, but, since we are trying to remove the density perturbations and not the 
actual noise, we can not reduce the measurement uncertainties. These uncertainties in $C_{\ell}^{TE}$ are then entirely determined by 
the noise in the original signal.

We have shown that the TE power spectrum due to PGWs is negative on large scales, hence a test determining whether 
the Wiener filtered power spectrum is negative or not is a probe of PGWs.

There are three different statistical tests we use to see if we can measure a negative TE power spectrum. The first test is a Monte Carlo 
simulation to determine signal-to-noise 
ratio, $S/N$ (Section \ref{montecarlo}). The other two tests are standard non-parametric statistical tests: the sign test 
(Section \ref{signintro}) and the Wilcoxon rank sum test (\cite{wilcoxonrst}) (Section \ref{wilcoxonintro}).

%\section{Monte Carlo Simulation} \label{simul}

For all of our tests, we calculate a random variable. If the data satisfies the hypothesis that $r=0$, we can calculate the mean and 
uncertainty in the variables. If 
we make one realization of data, the random variable is determined from its distribution. Because we are not using any real 
observational data, we must run a Monte 
Carlo simulation to reduce the risk of randomly getting a value for the variable taken from the outlying area of its 
distribution. To do this, the filtered 
multipoles, $C_{\ell,filt}^{TE}$, are randomly chosen from a gaussian distribution with mean $C_{\ell,t}^{TE}$ and standard 
deviation $\Delta C_{\ell}^{TE}$ where

\begin{eqnarray}
	\qquad (\Delta C_{\ell}^{TE})^2 &=& \frac{1}{(2\ell+1)f_{sky}} \Bigg( (C_{\ell}^{TE})^2 + \nonumber \\
	&& (C_{\ell}^{TT} + N_{\ell}^{TT})(C_{\ell}^{EE} + N_{\ell}^{EE}) \Bigg)
\end{eqnarray}
(see, for example, \cite{dodelsonbook}),
the variable $f_{sky}$ refers to the fraction of the sky covered by observations
and $N_{\ell}$ is the effective power spectrum of the instrumental noise (see \cite{dodelsonbook} for details on how $N_{\ell}$ is related to actual 
instrumental noise). 

Our determination of $C_{\ell,t}^{TE}$ is dependent on $\ell$. However, for two of our tests we ignore the value of $\ell$ in the calculation of 
the random variable. We assume that the calculated random variable is gaussian. In order for this to work, the random 
variable must be calculated from gaussian variables. The errors on the multipoles for the ``ideal'' toy experiment are large enough so that we 
can assume the multipoles are taken from a single distribution and not from a distribution that depends on $\ell$.

%Fig. \ref{Clfig} shows the combined distribution of the multipoles for the ``ideal'' toy experiment 
%(see Section \ref{resul}). To do this, for every $\ell$ we determine $10,000$ different 
%$C_{\ell,filt}^{TE}$ values. We then 
%combine all the different values, for every $\ell$, into one large distribution. Fig. \ref{Clfig} shows the errors on the multipoles are large 
%enough so that for our statistical tests we can assume the multipoles are taken from a single distribution and not from a distribution 
%that depends on $\ell$.
%
%\begin{figure}
%	\begin{center}
%		\includegraphics[totalheight=0.25\textheight]{ClDist.eps}
%		\\
%		\caption{\small  This plot shows the combined distribution of all the $C_{\ell,t}^{TE}$. $10,000$ values for every $\ell$ in the range 
%		$2 \le \ell \le 53$ were used in this histogram}
%		\label{Clfig}
%	\end{center}
%\end{figure}

\subsection{Monte Carlo S/N Test} \label{montecarlo}

For this test, the random variable we calculate, $S/N$, is defined as 

\begin{equation}
	S/N = \sum_{\ell=2}^{53} \frac{C_{\ell,t}^{TE}}{\Delta C_{\ell}^{TE}}.
\end{equation}
The reason why the sum in this equation is taken in the range $2<\ell<53$ is because only in this range $sgn(C_{\ell}^{TE}(\text{scalar})) 
= -sgn(C_{\ell}^{TE}(\text{tensor}))$. In other words, if we include higher multipoles we confront with a danger of a 
false detection, because the total TE power spectrum is negative for $\ell > 53$. 

The value of $S/N$ is gaussian distributed because it is a sum of many modes of squares of 
gaussian distributed values, $C_{\ell} = a_{\ell,m}^2$. We approximate each $C_{\ell}^{TE}$ as being gaussian distributed for the 
purpose of this paper. For each set of parameters we run this simulation one million times to 
determine the mean, $\left \langle S/N \right \rangle$, and 
standard deviation, $\sigma_{S/N}$. The mean of this distribution is determined by the preassumed value of $r$, while the standard deviation 
is determined by parameters of the experiment and gives the confidence level of detection. We run such Monte Carlo simulations for 
different values of $r$ to determine in what range 
of $r$ we can detect PGWs. When then using real observational data, we can 
compare the actual value of $S/N$ with the results of Monte Carlo simulations to infer the likelihood, as function of $r$,
which determines the probability that $r \not= 0$, or that PGWs exist at detectable levels.

\subsection{Sign Test} \label{signintro}

The sign test is a test of compatability of observational data with the hypothesis that $r=0$. If we do have $r=0$, 
then $C_{\ell,filt}^{TE}$ will be equally distributed around zero. Application of this test to the filtered data is very simple. 
In practice, all observational data are distributed 
between several bins and the averaging of the signal is produced in each bin separately. Let $N_{bins}$ be the number of such bins. 
The sign test actually gives the 
probability that in $N_-$ bins the average is negative and 
in $N_+=N_{bins}-N_-$ it is positive, if $r=0$. 
This probability, $P$, is given by the binomial distribution

\begin{equation}
	P(N_+) = \left( \begin{array}{c} N_{bins} \\ N_+ \end{array} \right) 0.5^{N_{bins}} = \frac{N_{bins}!}{N_+ ! N_- !} 0.5^{N_{bins}}
\end{equation}

The probability that the hypothesis $r=0$ is wrong is

\begin{equation}
	P(r\not=0) \approx 1 - 2\sum_{i=0}^{N_+} P(i)
\end{equation}
The value $\sum_{i=0}^{N_+} P(i)$ is the probability that we would get $\le N_+$ positive values given $r=0$. This is the same as the 
probability of getting $\le N_+$ negative values given $r=0$. Therefore our confidence that $r\not=0$ is just $100\%$ minus the 
sum of the probabilities describe above (the probability that the $N_+$ is closer to the mean, $N_{bins} / 2$, if $r=0$). This equation 
only makes sense if $N_+ < N_{bins}/2$, since that is required for $r > 0$. 
If $N_+ > N_{bins}/2$, that would imply $r < 0$, 
which is not physical. We would have to interpret the result as a random realization of $r \ge 0$, with the most likely result of $r = 0$. Therefore 
we would not be able to say $r \not= 0$ with any confidence.

Let us consider the following example: we put all measurements of $C_{\ell}^{TE}$ into $11$ bins and in three of them the average is 
positive. In this example, the 
probability that the hypothesis $r=0$ is wrong is equal to $89\%$.
 
One possible drawback of this method is that it does not take into account any measure of the signal-to-noise ratio of individual measurements. As we 
show in Section \ref{comptests}, it is possible to have 
two completely different sets of data with the same probability of having $r=0$. This test is also unable to make any prediction as to the value of $r$, 
only that it differs from zero.

\subsection{Wilcoxon Rank Sum Test} \label{wilcoxonintro}

This statistical test deals with two sets of data. The first set of data is taken from a real experiment which measures $C_{\ell}^{TE}$ with 
some unknown $r$. The second set of data is generated by 
Monte Carlo simulations (see Section \ref{montecarlo}) with $r=0$. The objective of the Wilcoxon rank sum test is to give the probability 
that the hypothesis $r=0$ is wrong (\cite{wilcoxonrst}). 

First, we choose some random variable $U$, whose probability distribution is known if $r=0$. For that, let us combine all data from first 
set with $n_1$ multipoles and second set with 
$n_2$ multipoles into one large data set, which obviously contains $n_1+n_2$ multipoles. Then, we rank all multipoles in the large data 
set from $1$ to $n_1+n_2$ 
according to their amplitude (rank $1$ for the smallest and rank $n_1+n_2$ for the largest). Now, the variables $R_1$ and $R_2$ are 
defined as the sum of the ranks for the first 
original data set and the second original data set, correspondingly. 
Finally, the variable $U$, is

\begin{eqnarray}
	\qquad \qquad \qquad U &=& \min(U_1,U_2) \text{, where} \nonumber \\
	U_i &=& R_i - n_i ( n_i + 1)/2 \text{, \qquad $i=1$,$2$}
\end{eqnarray}
If all multipoles of the first data set are larger than all multipoles of the second data set, then $U_1 = n_1 n_2$ and $U_2=0$. It is 
not difficult to show that $U_1 + U_2 = n_1 n_2$. 
If both sets of measurements have no evidence for PGWs, $\left \langle U_1 \right \rangle = \left \langle U_2 \right \rangle$. It is also simple
to see that $U_1 + U_2 = n_1 n_2$.

It is important to emphasize that the ranks of multipoles are random variables because 
all multipoles themselves are random variables, hence $U_1$, $U_2$, and $U$ are random variables. If $n_1 + n_2$ is large, the distribution 
of $U$ can be approximated as a gaussian 
with a known mean and standard deviation. In this approximation we have 

\begin{eqnarray}
	\qquad \qquad \qquad m_U &=& n_1 n_2 / 2 \\
	\sigma_U &=& \sqrt{\frac{n_1 n_2 (n_1 + n_2 + 1)}{12}}
\end{eqnarray}
In some cases, instead of $U$, the variables $R_1$ or $R_2$ are used. The reason $U$ is used here is because $m_U$ is symmetric 
in the data sets. If $r=0$ in both sets 
of data, then the distributions of $U_1$ and $U_2$ are the same, no matter what $n_1$ and $n_2$ are. The 
distributions of $R_1$ and $R_2$ would be the same only if $n_1 = n_2$. The probability that the first data set corresponds to $r \not= 0$ 
obtained from the test in which 
$R_1$ or $R_2$ is used is the same as if $U$ is used. 

Since this test requires Monte Carlo simulations for the second set of data, we ran this test many times for many different data sets 
to get an accurate mean value for $U$. 

To reject the hypothesis $r=0$ means to detect PGWs. Using the Wilcoxon rank sum test the allowable value of $r$ is determined, if
instead of comparing with simulated data with $r=0$, we compare with simulated data with $r=r_0 \not= 0$. 
In order to get a range of allowable values for $r$, we need to run multiple Monte 
Carlo simulations with multiple values for 
$r_0$. This is where the assumption that the $C_{\ell,t}^{TE}$ are from a random distribution that is independent of $\ell$ is 
used. This implies that the ranks are random 
variables. If the errors on the $C_{\ell,t}^{TE}$ are small enough, then the ranks will be predetermined. Therefore, our assumption 
about the distribution of $U$ will not be true and 
the test would have to be modified. Fortunately, this is not the case for even an experiment only limited by cosmic variance.

To illustrate how this test works, let us consider the following example. Assume there are $4$ multipoles in the first set of data and 
consider that $r=0.3$ is the correct value. There are also 
$4$ multipoles in the second set of data (which for sure corresponds to $r=0$). All quantities below are expressed in $\mu$K$^2$. 
The value for the first data set are
$C_{10}^{TE} = -0.005$, $C_{20}^{TE} = 0.02$, $C_{30}^{TE} = -0.015$, and $C_{40}^{TE} = -0.01$. The values for the second data set are 
$C_{10}^{TE} = 0.03$, $C_{20}^{TE} = 0.003$, $C_{30}^{TE} = -0.02$, and $C_{40}^{TE} = -0.003$. A ranking of multipoles gives the ordering from lowest 
to highest, with $1$ referring to the first data set and $2$ referring to the second data set, as $21112212$. This results in 
$R_1 = 2 + 3 + 4 + 7 = 16$, $U_1 = 16 - 10 = 6$, and 
$U_2 = 16 - 6 = 10$. Therefore $U=\min\{10,6\}=6$. For $n_1 = n_2 = 4$, to reject the hypothesis that $r=0$ at $95\%$ confidence 
level, $U_1$ should be less than one (see, for example, \cite{lehmannbook}). 
In this example, since $U_1 = 6 > 1$, the first set of data cannot be considered as a detection of PGWs.

\subsection{Comparison of Tests} \label{comptests}

The $S/N$ test is greatly affected by outlying measurements. A measurement of one large negative multipole could falsely imply a 
detection. Both the sign test and the 
Wilcoxon rank sum test are not affected by individual outlying measurements. In the sign test, the value of individual measurements 
is irrelevant, because the test is 
sensitive only to the sign of individual measurements. The Wilcoxon rank sum test is affected by outliers, but considerably less than 
the $S/N$ test. If the outlier 
is larger (or smaller) than every other multipole, its rank does not depend on its particular value.

If we have two completely different sets of data, the main disadvantage of the sign test, as mentioned in Section \ref{signintro}, is 
that it could give the same result, while for the two other tests the 
chance to obtain the same value of $r$ is negligible. For example, one set of data, consisting of $4$ small negative multipoles and 
$4$ large positive multipoles, gives 
the same result as another set of data, consisting 
of $4$ large negative multipoles and $4$ small positive multipoles. The $S/N$ test gives two very different values of $S/N$ for these 
two sets of data. We can also use the 
Wilcoxon rank sum test to compare these two sets of data. In this case $U = 16 = \frac{1}{2}m_U$, which corresponds to a confidence 
level of hypothesis that $r=0$ of less than $10\%$.

With observational data, the sign test can be applied and does not require any Monte Carlo simulations (which 
could be considered as an advantage of this test). The $S/N$ test requires Monte Carlo simulations, but only for the 
distribution of the random variable $S/N$. The Wilcoxon rank sum test requires large Monte Carlo simulations and combines the data sets 
generated by these simulations with observational data. In other words, Monte Carlo simulations are absolutely necessary after 
obtaining observational data, which may be considered a disadvantage of this test.
Thus, each of the three tests has advantages and disadvantages, suggesting that the best way to work out observational 
data is to apply all these three tests.

\section{Discussion and Results} \label{resul}

\cite{baskaran06} used equal amplitudes of scalar and tensor perturbations to sharpen the discussion in their plots. They defined 
the tensor-to-scalar ratio, $R$, as the ratio of the 
temperature quadrupoles, $R \equiv C_{2,t}^{TT}/C_{2,s}^{TT}$ and set $R = 1$. Using 
standard WMAP3 cosmological parameters (\cite{Spergel2007WMAP}), the definition of tensor-to-scalar ratio, $r$, used in this paper
is approximately twice as large as their definition of $R$. The exact relationship between $r$ and $R$ will depend on the cosmological 
parameters used. This means that $R=1$ is equivalent to $r \approx 2$, which has currently been strongly ruled out by 
WMAP in combination with previous experiments (\cite{Spergel2007WMAP}). We need to see 
if this method can detect a value of $r$ that is currently within the limits.
We assume that there is no foreground contamination. In reality foregrounds affect the measured location of $\ell_0$ (we will 
consider the effects of foregrounds on $\ell_0$ elsewhere). For 
the experiments that do not 
observe the full sky, correlations between multipoles must be taken into account. The multipoles are binned together of such width that the correlations 
between the bins are sufficiently small.

%%%%%%%%%%%%%%%%%%%%Experiments section
Two different toy experiments, along with the two satellite experiments WMAP and Planck, are considered to constrain $r$. The first toy
experiment is a full sky experiement. It is idealized in two aspects. The first idealization is that we can take measurements 
over the full sky while the second idealization is 
that we assume there is no detector noise. The only uncertainty is due to cosmic variance. Such experiment represents the 
best limit to which the gravitational waves can be detected by the CMB TE correlation. This toy experiment is close to a space-based 
experiment with access to the full sky. It is similar to what the Beyond Einstein inflation probe would be able to detect. 
This toy experiment will be hereafter referred to as the ideal experiment. 
The second toy experiment is a more realistic one.  In this experiment, measurements are on $3\%$ of the sky, the frequency is $100$ GHz, and the 
duration of the experiment is $3$ years. The 
noise in each detector of the $50$ polarization sensitive bolometer pairs can be described by their noise equivalent temperature (NET) of 
$450$ $\mu K \sqrt{s}$. The detectors' beam 
profile is assumed to be gaussian and and it is described by the width at half of the maximum sensitivity, abbreviated as FWHM of $0.85^{\circ}$.

This second toy experiment is similar to current ground-based experiments and the constraints from this experiment represent those 
that can and will be obtained in the next several years. This will be referred to as the realistic experiment.

The predicted errors for Planck are based on using the $100$ GHz, $143$ GHz, and $217$ GHz channel in the High Frequency Instrument (HFI). The numbers are 
gotten from the Planck science case, the ``bluebook''\footnote{http://www.rssd.esa.int/index.php?project=Planck}. The WMAP noise was obtained by $3$ years 
of observations of the Q-band, V-band, and W-band detectors. 

%%%%%%%%%%%%%%%%%%%End of experiments section

\subsection{Ideal Experiment} \label{toy1res}

\begin{figure}
	\begin{center}
		\includegraphics[totalheight=0.25\textheight]{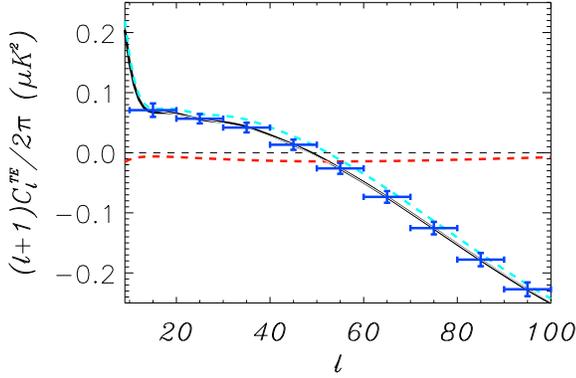}
		\\
		\caption{\small  The black line is the total TE mode with a $r=0.3$. The red line is the contribution from PGWs 
		only while the light blue line is the contribution from the density perturbations. Blue is the error bars for the ideal experiment 
		binned in intervals of $\Delta \ell = 10$.}
		\label{exp1r3}
	\end{center}
\end{figure}

A plot of TE for $r=0.3$ and $n_t=0.0$ with error bars for $\ell$ binned in bins of width $\Delta \ell = 10$ is shown 
in Fig. \ref{exp1r3}. This figure separately shows the contribution to the TE mode of density perturbations, contribution of PGWs 
with $r=0.3$, when the TE power spectrum due to density perturbations is approximately five times 
larger than the power spectrum due to PGWs at $\ell < \ell_0$.

\begin{figure}
	\begin{center}
    	\includegraphics[totalheight=0.25\textheight]{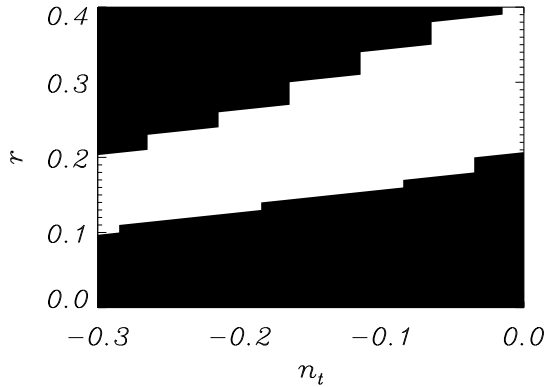}
    	\\
    	\caption{\small This is a plot of the allowed $r$ and $n_t$ for the $1\sigma$ region of $\ell_0$ for the ideal experiment. The white is 
		the $1\sigma$ region while the black is the forbidden region}
		\label{exp1detect}
	\end{center}
\end{figure}

The Monte Carlo simulation for the calculation of $\ell_0$, with an input model of $r=0.3$ and $n_t=0$, results in the value of 
$\ell_0 \approx 49$ with an uncertainty of 
$\Delta \ell_0 \approx 1.3$. A contour plot of the limits on the resulting measurement of 
$r$ is shown in Fig. \ref{exp1detect}. The white is the allowed region for $r$ and $n_t$ that falls within the $1\sigma$ errors of $\ell_0$. The black 
is the region forbidden with $68\%$ confidence. If $n_t = 0$, then we measure $r \approx 0.3 \pm 0.1$. If we consider the inflationary consistency 
relation, $n_t = -r/8$ (\cite{peiris03}), we then get the constraint $r=0.3^{+0.09}_{-0.1}$. The uncertainty is smaller, but not by much. We predict 
a $3\sigma$ detection of PGWs by the zero multipole method. 

\begin{figure}
	\begin{center}
    	\includegraphics[totalheight=0.25\textheight]{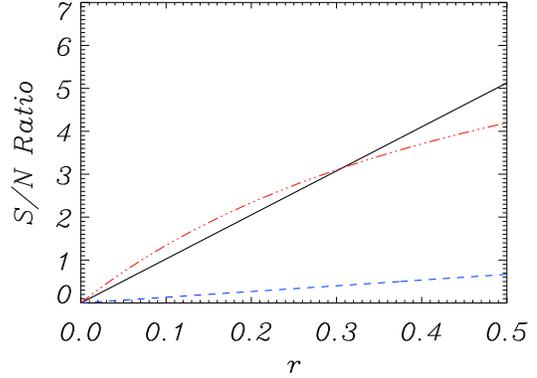}
    	\\
    	\caption{\small The signal-to-noise ratio for the zero multipole method are shown as the solid black, for ideal experiment, 
		and dashed blue, for realistic experiment, lines. The 
		signal-to-noise ratio for realistic measurements of the BB power spectrum is shown as the dash-dot red curve. For all curves $n_t=0$.}
		\label{ell0detect}
	\end{center}
\end{figure}

The detectability of $\ell_0$ using the ideal experiment is shown in Fig. \ref{ell0detect}. For $n_t=0$ the effective number of $\sigma$ detection is 
$\sigma \approx 10r$. We make this approximation by determining the detectability for several values of $r$ and then approximating 
a line. For comparison the results are also shown for the zero-multipole method with the realistic experiment and for measurements of the 
BB power spectrum with the realistic experiment described above. We assume we 
can make measurements over a range of $60$ multipoles for BB measurements.

%\begin{figure}
%	\begin{center}
%		\includegraphics[totalheight=0.25\textheight]{ClErrCV.eps}
%		\\
%		\caption{\small  The black line is the TE mode due to PGWs with $r=0.3$. Red is the error bars for the ideal experiment 
%		calculated from the total TE power spectrum 
%		binned in intervals of $\Delta \ell = 10$.}
%		\label{exp1r3}
%	\end{center}
%\end{figure}

\begin{figure}
	\begin{center}
    	\includegraphics[totalheight=0.25\textheight]{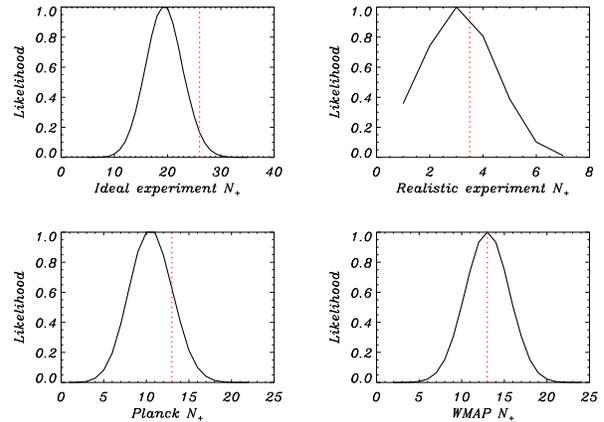}
    	\\
    	\caption{\small This is a plot of the distribution of the number of positive multipoles for the Monte Carlo simulation for 
		the ideal experiment (upper left), 
		the realistic experiment (upper right), Planck (lower left), and WMAP (lower right). The dotted red line shows where $N_+ = \frac{1}{2}N_{bins}$}
		\label{signdist}
    \end{center}
\end{figure}

The Monte Carlo simulation for the Wiener filtering gives an average of $19$ measured TE power spectrum multipoles greater than zero out of a 
total of $52$ independent multipoles. If the null hypothesis was true, 
the sign test would indicate there is a $3.5\%$ chance of measuring $\le 19$ positive multipoles. This is equivalent to a
$\approx 1.8\sigma$ detection. 
A plot of the distribution of the number of positive multipoles is shown in the upper panel plot of Fig. \ref{signdist}. There is an $81\%$ chance for the 
observed $N_+$ to give a $1\sigma$ detection of PGWs. 

\begin{figure}
	\begin{center}
    	\includegraphics[totalheight=0.25\textheight]{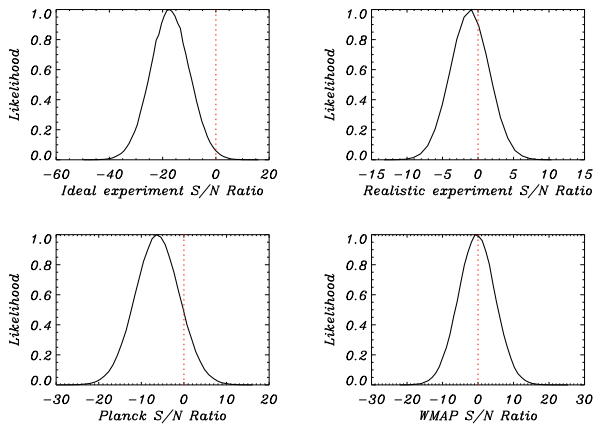}
    	\\
    	\caption{\small The $S/N$ statistic distribution for the ideal experiment (upper left), realistic experiment (upper right), 
		Planck (lower left), and WMAP (lower right). The dotted red line shows where $S/N=0$.}
		\label{SNPlot}
    \end{center}
\end{figure}

\begin{figure}
	\begin{center}
    	\includegraphics[totalheight=0.25\textheight]{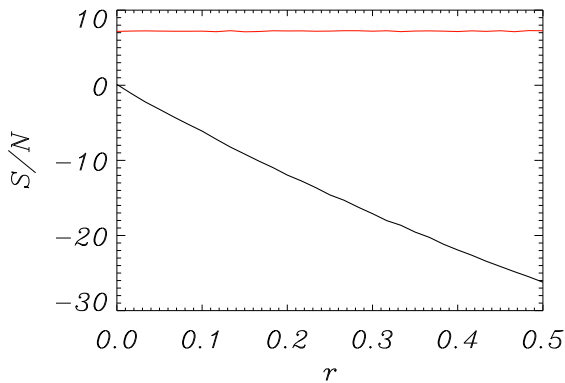}
    	\\
    	\caption{\small This is a plot of $\left \langle S/N \right \rangle$ and $\sigma_{S/N}$ as a function of $r$ for the ideal 
		experiment. The black line is $\langle S/N \rangle$ 
		and the red line is $\sigma_{S/N}$.}
		\label{SNmeanSD}
    \end{center}
\end{figure}

The $S/N$ test gives a mean value of $S/N = -17.1$ and standard deviation of 
$7.21$. The upper left panel in Fig. \ref{SNPlot} shows the 
distribution of the $S/N$ values for the Monte Carlo simulation with $r=0.3$. If $r=0.3$ we would have a $0.8\%$ probability of 
the measured $S/N > 0$. This negative value signifies that a non-zero tensor-to-scalar ratio produced an anti-correlation. We can assume 
that the standard deviation would be the same if the mean of $S/N$ was $0$ (equivalent 
to $r=0.0$), because it is equivalent to adding a constant value to every measured value (and hence adding a constant to $S/N$ which 
would not change the error). Therefore, if $r=0$, the 
probability of getting $S/N < -17.4$ is $0.8\%$, and hence we have a $99\%$ chance that $r \not= 0$. A plot of 
$\left \langle S/N \right \rangle$ and $\sigma_{S/N}$ 
as a function of $r$ is shown in Fig. \ref{SNmeanSD}. As can be seen from the plot, we can predict a value of $r$ for any value of 
$S/N$. The value of $\sigma_{S/N}$ is a 
relatively constant function of $r$ and so our prediction about the distribution of $S/N$ for different value of $r$ is a good 
approximation to the true distribution.

\begin{figure}
	\begin{center}
    	\includegraphics[totalheight=0.25\textheight]{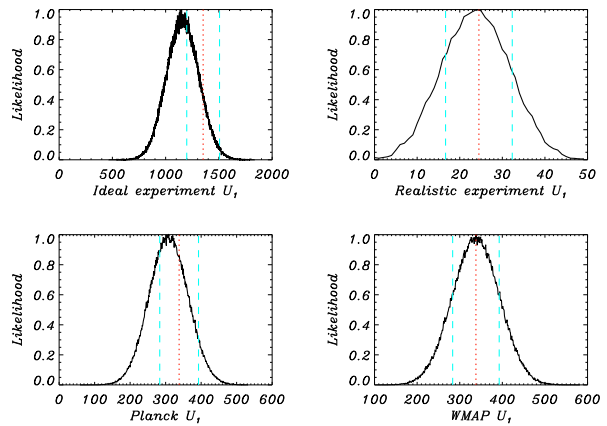}
    	\\
    	\caption{\small This is the plot of the distribution of $U$ for the ideal experiment (upper left), realistic experiment 
		(upper right), Planck (lower left), and 
		WMAP (lower right). The red dotted line is the value for $m_U$ and the light blue dashed lines enclose the $1\sigma$ 
		region for $U$ assuming the hypothesis that $r=0$}
		\label{WilcoxonPlot}
    \end{center}
\end{figure}

The Wilcoxon rank sum test gives $U_{avg}-m_U = -1.23\sigma_U$. The variable $U_{avg}$ is the 
mean value for $U$ in the Monte Carlo simulations described earlier. The 
values $m_U$ and $\sigma_U$ are given in Section \ref{wilcoxonintro}. The distribution of $U$ for the Monte Carlo simulations with 
$r=0.3$ is shown in Fig. \ref{WilcoxonPlot}. The standard 
deviation of the distribution of measured $U$ is the same as the standard deviation of the distribution of $U$ assuming the hypothesis 
that $r=0$. The only difference between 
the distributions is that $m_U$ is shifted by a constant value. Therefore, there is a $22\%$ chance 
that $U-m_U < -2\sigma_U$. There is also a $40\%$ chance that we measure $U-m_U < -1\sigma_U$, and are not even able to make a 
$1\sigma$ detection of PGWs.

\begin{figure}
	\begin{center}
    	\includegraphics[totalheight=0.25\textheight]{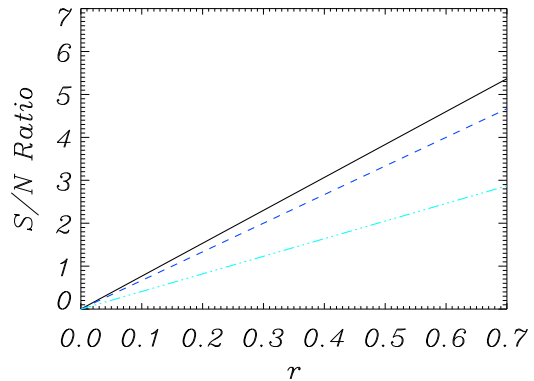}
    	\\
    	\caption{\small This is the plot of the signal-to-noise ratio (number of $\sigma$s) for different values of $r$ for the 
		three different tests. The black line is the $S/N$ test, the dashed dark blue 
		line is the sign test, and the dotted-dashed light blue line is the Wilcoxon rank sum test}
		\label{Detectability}
    \end{center}
\end{figure}

A comparison of the three tests is shown in Fig. \ref{Detectability}. This is obtained by simulated with with several values of $r$ 
and then interpolating between them. A $2\sigma$ detection is 
obtained for $r=0.26$ ($S/N$ test), $r=0.3$ (sign test), and $r=0.5$ (Wilcoxon rank sum test), highlighting its intended use as 
a monitor of a false positive detection for large $r$.

\subsection{Realistic Ground Based Experiment} \label{toy2res}

\begin{figure}
    \begin{center}
    	\includegraphics[totalheight=0.25\textheight]{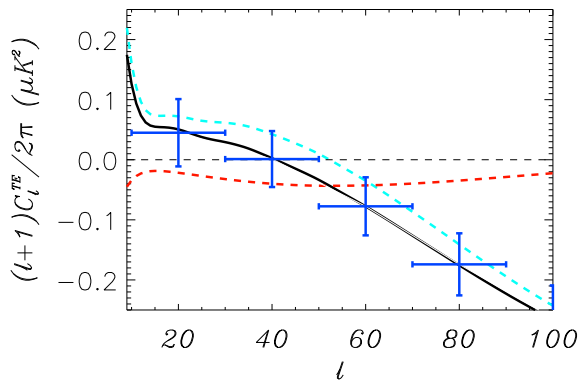}
    	\\
		\caption{\small This figure is the same as Fig. \ref{exp1r3} but presents for the realistic experiment 
		with $r=0.9$.}
		\label{exp2r9}
	\end{center}
\end{figure}

A plot of the error bars for the realistic experiment is shown in Fig. \ref{exp2r9} with $r=0.9$. Observations on an incomplete sky 
require the multipoles to be binned in sizes of $\Delta \ell = 10$. This experiment has much 
larger error bars than the ideal experiment and it is not able to detect low values of $r$ with the 
TE cross-correlation only. Plots of the TE power spectrum due to density 
perturbations and PGWs are shown in Fig. \ref{exp2r9} along with the combined TE power spectrum.

\begin{figure}
    \begin{center}
    	\includegraphics[totalheight=0.25\textheight]{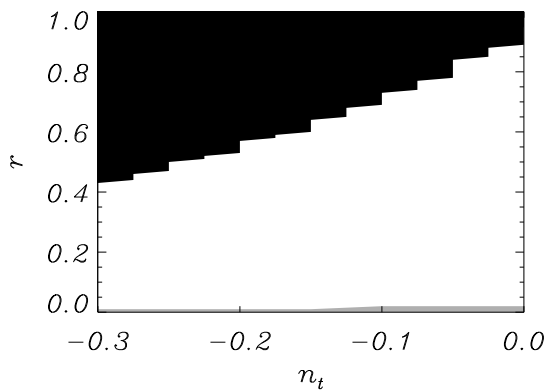}
    	\\
    	\caption{\small An upper limit on $r$ due to the realistic experiment with $\Delta l_0 = 10$}
		\label{figuldell10}
	\end{center}
\end{figure}

For this experiment, the constraints on measuring $\ell_0$ are significantly larger than those for the ideal experiment. The $1\sigma$ 
uncertainty on $\ell_0$ is $\Delta \ell_0 \approx 10$. This corresponds to a limit of $r < 0.9$ with $68\%$ confidence. If we want a $2\sigma$ 
limit, then the constraint expands to $r \le 1.5$. If we assume the inflationary consistency relation, then this error on $\ell_0$ 
would correspond to a $1\sigma$ upper limit of 
about $r \lesssim 0.7$. Fig. \ref{figuldell10} shows the region of $r$ and $n_t$ allowed with $68\%$ confidence of $\ell_0$. 

As mentioned earlier, Fig. \ref{ell0detect} shows the signal-to-noise ratio of the zero multipole method for the realistic experiment. 
The measurements of the BB power spectrum are much more sensitive to PGWs and the sensitivity is roughly the same as in the ideal 
experiment. This shows that the zero multipole method is less sensitive to PGWs than measurements of the BB power spectrum.

%\begin{figure}
%    \begin{center}
%    	\includegraphics[totalheight=0.25\textheight]{ClErrBicep.eps}
%    	\\
%		\caption{\small This figure is the same as Fig. \ref{exp1r3} except for the realistic experiment}
%		\label{exp2r9}
%	\end{center}
%\end{figure}

The results for the Wiener filtering method are much worse than those for the ideal experiment for $r=0.3$. Since this experiment 
observes a small portion of the 
sky, the multipoles are correlated and we must bin together to get reasonably uncorrelated measurements. For this experiment, we 
only have $7$ to $8$ uncorrelated multipoles, 
instead of $52$ uncorrelated multipoles in the case where the full sky is observed. Getting $7$ out of $8$ negative multipoles is 
a $3\%$ probability if there are no 
PGWs. For the Monte Carlo simulations of the realistic experiment, on average, half of measured multipoles are positive and half 
are negative. A plot of the distribution of the number of positive multipoles is shown in the upper right panel of Fig. \ref{signdist}. In this 
case, we cannot distinguish $r=0.3$ from $r=0.0$ with any significance. 

The $S/N$ test gives an average value of $S/N=-0.95$ with 
standard deviation of $2.64$. For the realistic toy experiment, the distribution of $S/N$ for $r=0.3$ is shown in the upper right panel of Fig. 
\ref{SNPlot}. In order to obtain $68\%$ confidence detection of 
PGWs, we must use $r \approx 0.7$. In this sense the TE test provides monitoring and insurance against false positive detection with $r > 0.7$, 
which could arise, for example, if foregrounds or other systematic effects arer improperly removed. 

The last statistical test, the Wilcoxon rank sum test, gives $U_{avg} - m_U = -0.20 \sigma_U$. The distribution of $U$ for $r=0.3$ is 
shown in the upper right panel of 
Fig. \ref{WilcoxonPlot}. This gives the weakest result in terms of the three tests for the Wiener filtered data. 
The realistic experiment will not be able to constrain $r < 0.3$ using the TE cross correlation power spectrum. Its 
limit is closer to $r < 0.7-0.9$ at only $68\%$ confidence depending on the test used. For a higher confidence in a detection of PGWs, 
the value of $r$ would need to be much higher. Since the observed distribution of 
$U$ corresponds almost exactly to the simulated distribution of $U$ under the assumption that $r=0$, therefore we have a $16\%$ chance of 
measuring $U-m_U < -1\sigma_U$. 

\subsection{WMAP} \label{wmapres}

A constraint on $r$ using a measurement of $\ell_0$ for WMAP is almost impossible. Using error bars consistent with WMAP noise, 
we get $\Delta \ell_0 \approx 15$ for an input of $r=0.3$ and $n_t=0$. The published 
results of WMAP give limits of $r<0.3$ so adding this method to the WMAP results would not change constraints significantly. In fact, 
using the real WMAP data\footnote{http://lambda.gsfc.nasa.gov/} we get $\ell_0 \approx 48$. With an uncertainty of $\Delta \ell_0 \approx 15$, 
the probability of getting a value farther away from $\ell_0 = 53$ is larger than $50\%$, so we cannot detect primordial gravitational 
waves in the published WMAP data using the zero multipole method.

The results of the Wiener filtering showed that the WMAP cannot make a detection 
of gravitational waves using the TE cross correlation power spectrum alone. As with the two toy experiments, the result of the 
scalar and tensor separation was similar.
The Monte Carlo simulation gave on average gave $13$ positive multipoles out of a total of $26$ uncorrelated multipoles. We 
would get the same result if the input data had 
$r=0.0$ so we cannot detect PGWs with WMAP using only the TE power spectrum. A plot of the distribution of the number of 
positive multipoles is shown in the lower right panel of 
Fig. \ref{signdist}. As can be seen, this distribution of $N_+$ for WMAP noise and $r=0.3$ is simply the distribution for $r=0$.  

For WMAP, the S/N test gives the value of $S/N=-0.02$ with a standard deviation of 
$5.09$. The distribution is shown in the lower right panel of Fig. \ref{SNPlot}. The distribution is centered around $S/N=0$ 
so there is no chance of using this 
test to detect PGWs in WMAP's TE power spectrum. The probability of getting a $1\sigma$ or $2\sigma$ detection is the same 
probability that we would randomly get a detection if there are no PGWs.

The rank sum test gives a value of $U_{avg}-m_U = -0.004\sigma_U$, 
which is implies no ability to distinguish WMAP's observed TE data from a data set with no PGWs. 
A plot of the distribution of $U$ for WMAP error bars is shown in the lower right panel of Fig. \ref{WilcoxonPlot}. We reach the  
same conclusion for WMAP noise as for 
the realistic experiment. There is only a $16\%$ chance that we can measure $U-m_U < -1\sigma_U$ and make a $1\sigma$ detection of $r=0.3$

The published WMAP results show an anti-correlation of TE power spectrum at large scales. Unfortunately this is not a detection of PGWs as 
theorized in \cite{baskaran06}. The contribution to the TE power spectrum due to PGWs only changes sign once for $\ell \lesssim 90$. If a claimed 
evidence for 
gravitational waves is to be believed, then the TE power spectrum would have to change sign three times for $\ell \lesssim 60$. In fact, other than 
the two anticorrelations at low $\ell$, the rest of 
the multipoles, up to $\ell=53$, are consistent with $r=0$. None of the described tests applied to the current WMAP data will give any detection of PGWs.

\subsection{Planck} \label{planckres}

The uncertainty in $\ell_0$ is much better for Planck than for the realistic experiment and about twice as large for the ideal experiment. 
The Monte Carlo simulations resulted in $\Delta \ell_0 \approx 3.75$ for an input TE power spectrum with $r=0.3$ and $n_t = 0$. 
This results in $\approx 68\%$ confidence that $r \not= 0$, under the assumption that $n_t=0$.

The sign test gives on average $10$ positive measurements of the TE power spectrum out of a total of $26$ uncorrelated multipoles. 
There is a $16\%$ chance of getting $\le 10$ positive 
multipoles if $r=0$. A plot of the distribution of the number of positive multipoles for Planck is shown in the lower left panel of 
Fig. \ref{signdist}. There is a $50\%$ 
chance that we will measure $N_+ < 10$ and hence have a $1\sigma$ detection of $r=0.3$.

The $S/N$ test gives a value of $S/N=-6.24$ with 
a standard deviation of $5.09$. There is only a $10\%$ chance that the $S/N$ test results in a value of $S/N$ larger than zero, 
if $r=0.3$, and a $10\%$ chance getting $S/N < -3.12$ if $r=0$. 
This is close to a $90\%$ probability of detection. The distribution of the $S/N$ variable is shown in lower left panel of Fig. \ref{SNPlot}. 

Again, the rank sum test gives the lowest confidence result with a value of $U_{avg}-m_U = -0.66\sigma_U$. A plot of the distribution 
of $U$ is shown in the lower 
left panel of Fig. \ref{WilcoxonPlot}. There is a $37\%$ probability that we will measure $U-m_U < -1\sigma_U$ and a $9\%$ probability that we 
measure $U-m_U < -2\sigma_U$ for Planck.

\section{Comparison of Measurements of the TE Power Spectrum with the BB Power Spectrum} \label{comptebb}

As mentioned earlier, it was originally suggested that it might be easier to detect PGWs using the TE power spectrum instead 
of the BB power spectrum. For both methods, this 
turned out not to be true. The reason for this is because we are trying to measure the TE power spectrum at the place where the 
signal is lowest ($C_{\ell}^{TE} = 0$). In measurements of the 
BB power spectrum, if we neglect instrumental noise, the signal decreases with a decrease in $r$ and so does the 
cosmic variance limited uncertainty. This is not the case for the TE power spectrum. The uncertainty in 
the measurement of the TE power spectrum due to PGWs is determined by the total TE, TT, and EE power spectra. When the TE power spectrum 
goes to zero, the TT and EE power spectrum do not approach zero (in fact, they increase as we approach to $\ell_0$). We therefore have 
a low signal-to-noise ratio around $\ell_0$ making it very hard to 
detect PGWs using the zero multipole method. Below we give simple summarizing arguments why the same is true for the Wiener filtering 
of the TE power spectrum

If $N_{\ell} \ll C_{\ell}^{BB}$, the signal-to-noise ratio for the BB power spectrum is

\begin{equation}
	(S/N)_{BB} =  \frac{C_{\ell}^{BB}}{\Delta C_{\ell}^{BB}} = \gamma \frac{C_{\ell}^{BB}}{C_{\ell}^{BB} + N_{\ell}} \approx \gamma,
\end{equation}
where

\begin{equation}
	\gamma =  \sqrt{\frac{(2\ell+1)f_{sky}}{2}}
\end{equation}
If $N_{\ell} > C_{\ell}^{BB}$ then we will not be 
able to detect PGWs and a comparison with the TE power spectrum is not worthwhile. 

If $N_{\ell} \ll C_{\ell}^{EE}$ and $r < 1$, for the TE power spectrum, the signal-to-noise ratio is

\begin{eqnarray}
	(S/N)_{TE} &=& \frac{C_{\ell,t}^{TE}}{\Delta C_{\ell}^{TE}} \nonumber \\
	&=& \sqrt{2} \gamma \frac{C_{\ell,t}^{TE}}{\left[ (C_{\ell}^{TE})^2 + (C_{\ell}^{TT} + N_{\ell}/2)(C_{\ell}^{EE} + N_{\ell})\right]^{1/2}} \nonumber \\
	&\approx& \sqrt{2} \gamma  \frac{C_{\ell,t}^{TE}}{\left[ (C_{\ell}^{TE})^2 + C_{\ell}^{TT}C_{\ell}^{EE}\right]^{1/2}} \nonumber \\
	&\approx& \sqrt{2} \gamma \frac{r}{\alpha + \beta r}
\end{eqnarray}
where $\alpha$ and $\beta$ are

\begin{eqnarray}
	\alpha &=& \frac{\sqrt{(C_{\ell,s}^{TE})^2 + C_{\ell,s}^{TT} C_{\ell,s}^{EE}}}{D_{\ell}^{TE}}, \nonumber \\
	\beta &=&  \frac{2C_{\ell,s}^{TE} D_{\ell}^{TE} + D_{\ell}^{TT} C_{\ell,s}^{EE} + C_{\ell,s}^{TT} D_{\ell}^{EE}}{2 D_{\ell}^{TE} \alpha}, 
\end{eqnarray}
where
\begin{equation}
	D_{\ell}^{XY} = C_{\ell,t}^{XY} / r
\end{equation}
One can see that $\alpha$ and $\beta$ are on the order of unity. Therefore, the signal-to-noise ratio is approximated as

\begin{equation}
	(S/N)_{TE} = \sqrt{2} \gamma \frac{r}{\alpha+\beta r} \approx \sqrt{2} \gamma \frac{r}{\alpha}
\end{equation} 
In other words if $r < \alpha/\beta \sim 1$, BB measurements have the obvious advantage in comparison with the Wiener filtering of 
the TE power spectrum. Indeed if $r \lesssim 1$, 
$(S/N)_{BB} \sim \gamma$, while $(S/N)_{TE} \sim \gamma r < \gamma$. This is because in BB measurements, applying proper data 
analysis, we can entirely eliminate contributions of scalar perturbations to CMB polarization signal as well as to the uncertainties. 
For the perfect Wiener filtering of the TE power 
spectrum, we can eliminate the contribution of scalar perturbations to the signal only, but cannot eliminate their contribution to the uncertainties.

\section{Conclusion}

The measurement of where the TE cross correlation first changes sign can be used to detect or put constraints on PGWs. Such 
constraints are not as strong as the ones given by measurements of 
the BB power spectrum, however it is useful to have a supplementary method to detect PGWs. We have shown how well the TE 
mode can constrain the amount of PGWs from just a measurement of the angular scale where it first changes sign for two 
different toy experiments and two real satellite experiments. The absolute best limit with which we can measure $\ell_0$ only gives us less 
than a $3\sigma$ detection 
of the PGW component if $r=0.3$. The current confidence limits gives us $r < 0.3$ at $95\%$ confidence level. Current and future 
experiments are optimized to measure the BB power spectrum if $r \le 0.1$ even in the presence of foregrounds, which are not 
taken into account in this paper. Future satellite 
experiments should be able to detect $r<0.01$ which is $10$ times better than the sensitivity to $r$ than the result of the 
ideal experiment. If one neglects even cosmic variance, the discreteness of $\ell$ limits the 
calculation of $\ell_0$, and the sensitivity to $r$, to values considerably larger than $0.01$. The cosmic variance is largest 
at low $\ell$ and is proportional to the total power spectrum. 
Since the TE cross correlation has contributions from density perturbations the errors in the measured TE power spectrum make detecting 
deviations of $\ell_0$ from $53$ difficult, though they also provide insurance against a false detection or imperfect subtraction of instrumental 
and foreground systematic effects.

The other method described in this paper is one in which we filter out the signal due to density perturbations, leaving only the contribution to the 
TE power spectrum due to PGWs. We 
then test the resulting TE power spectrum to see if it is negative. 
Three different statistical tests were used to see if there was a significant detection of PGWs. The $S/N$ test can 
give a value for $r$ using a comparison with Monte Carlo simulations, while the Wilcoxon rank sum test can only give an allowable range for $r$. The 
sign test will only tell us if $r \not= 0$. 

Using the Wiener filtering method, we are unable to make as significant of a detection as using the zero multipole method. The best 
result was for the $S/N$ test which would 
give a $2.3\sigma$ detection of $r=0.3$. To detect PGWs on the level of $3\sigma$, the tensor-to-scalae ratio $r$ should be $r \ge 0.4$. The 
sign test would give $2\sigma$ detection for $r=0.3$ and a 
$3\sigma$ detection for $r=0.45$. The Wilcoxon ranked sum test gives only a $1.2\sigma$ 
detection for $r=0.3$ and a $3\sigma$ detection for $r=0.7$. Similar results were gotten for the other three experiments tested. Thus 
in the sense of potential to detect PGWs, the zero 
multipole method is the best, next best is the $S/N$ test, then the sign 
test, and the worst is the Wilcoxon ranked sum test. 

\cite{baskaran06} present illustrative examples in which high $r$ is consistent with measured TT, EE, and TE correlations. The 
value of $r$ is so high in these examples  
that if PGWs with such $r$ really existed, current BB experiments would already detect PGWs. 
All models predict that the TE cross correlation power spectrum change sign only once for $\ell < 100$. The fact WMAP cannot exclude 
several multipoles with $C_{\ell}^{TE} > 0$ in between multipoles of $C_{\ell}^{TE} < 0$ means that the TE cross correlation power 
spectrum either changes sign several times for $\ell < 100$ or there is some instrumental noise which causes some anticorrelation measurements. Using 
instrumental noise consistent with WMAP, our Monte Carlo simulations give $\Delta \ell_0 \approx 16$ and $\ell_0 > 40$, which means 
that there is no evidence of 
PGWs in the TE correlation power spectrum. 

\section*{Acknowledgments}

AGP would like to thank the Center for Astrophysics and Space Sciences at UCSD for hosting him 
while working on parts of this paper and Deepak Baskaran and Leonid Grishchuk for helpful discussions. NJM would like to thank the Astronomy Unit, 
School of Mathematical Science at Queen Mary, University of London for hosting 
him while working on parts of this paper. BGK gratefully acknowledges support from NSF PECASE Award AST-0548262. 
We acknowledge helpful comments on this manuscript by Kim Griest and Manoj Kaplinghat. We acknowledge 
using CAMB to calculate the power spectra in this work.

\label{lastpage}

\bibliographystyle{mn2e}
\bibliography{Inflation,Polarization}
\end{document}